\documentclass[aps,twocolumn,showkeys]{revtex4-1}
\usepackage{graphicx}
\usepackage{amssymb,amsfonts,amsmath}
\usepackage{epstopdf}
\usepackage{subfigure}

\newcommand{\remove}[1]{}

\begin{document}



\title{The Simple Rules of Social Contagion}

\author{Nathan O. Hodas and Kristina Lerman}
\affiliation{USC Information Sciences Institute\\ Marina del Rey, CA 90292}
\date{\today}


\begin{abstract}
 It is commonly believed that information spreads between individuals like a pathogen, with each exposure by an informed friend potentially resulting in a naive individual becoming infected.  However, empirical studies of social media suggest that individual response to repeated exposure to information is significantly more complex than the prediction of the pathogen model.
 As a proxy for intervention experiments, we compare user responses to multiple exposures on two different social media sites, Twitter and Digg. We show that the position of the exposing messages on the user-interface strongly affects social contagion.
 Accounting for this visibility significantly simplifies the dynamics of social contagion. The likelihood an individual will spread information increases monotonically with exposure, while explicit feedback about how many friends have previously spread it increases the likelihood of a response.  We apply our model to real-time forecasting of user behavior.
 \end{abstract}

\maketitle
People talk about ideas and information spreading between people like viruses, using phrases like ``social contagion,'' ``contagious ideas'' and ``viral content.''
This biological metaphor serves as a starting point for many analyses of information spread~\cite{newman02,Kempe03,Gruhl04,Anagnostopoulos08}.  However, recent works have observed that multiple exposures to a piece of information suppress an individual's response, thus suggesting that social contagion is more complex than originally thought~\cite{Romero:2011um,VerSteeg11}.  Our work clears the confusion and shows that there are important and surprising differences between the spread of information and the spread of disease,  stemming from human cognitive limitations for discovering information. Once we account for the information discovery process, social contagion is actually quite simple and people's responses can be accurately predicted.

From a theoretical perspective, the simplest and most widely studied model of social contagion is the independent cascade model (ICM), 
which also serves as the basis of the SIS and SIR models of biological epidemics~\cite{Hethcote00}. The ICM assumes that each exposure of a healthy (naive) person by an infected (informed) friend leads to an independent chance of information transmission. Therefore, the probability that a healthy individual becomes infected increases monotonically with the number of exposures, potentially causing a global epidemic involving a substantial fraction of the population~\cite{Castellano:2009ce,satorras2001}.  However, studies of information spread in social media have identified social behaviors that qualitatively differ from predictions of the ICM. For example, when measuring how people respond to their friends' use of certain memes or recommendations for news articles, repeated exposure initially increases infection probability, but eventually exposure appears to be inhibitory~\cite{Romero:2011um,VerSteeg11}. A number of explanations have been offered for this aberration, including complex contagion~\cite{Granovetter83, Watts02,Centola07b}. In complex contagion, the probability to adopt a behavior, or an idea, varies with the extent of exposure, suggesting that social phenomena may drive respnse and interact non-trivially with network structure~\cite{Centola:2010ww, GonzalezBailon:2011kz, Mason12}. Among other factors thought to affect social contagion are the novelty~\cite{Wu07} or persistence~\cite{Romero:2011um} of information, and competition with other information~\cite{Weng:2012dd}. The role of cognitive constraints in online social interactions has not been widely examined, although one study of Twitter demonstrated that people limit themselves to approximately 150 conversation partners~\cite{Goncalves:2011im}, a number similar to the bound on human social group size~\cite{Dunbar}.

In this Letter, we present novel techniques, originating from non-equilibrium statistical physics, to analyze user behavior data from two online social networks, Digg and Twitter.  Our approach enables us to separate the factors of social contagion that are attributable to the visibility of information (i.e., how easily it can be discovered in the user interface of each site) from the factors attributable to social influence. We reveal
that, after accounting for these factors, contagion becomes quite simple: each exposure increases the likelihood of a response, and social signals about the number of friends who have previously adopted the information (when such signals are provided by the web site) further amplify the probability of a response.  This result implies that people are much  more susceptible to social media campaigns than previously believed.  Finally, we show that we are able to forecast an individual's behavior in real-time on both sites.

\section{Data Sources}
To compare how visibility and social factors contribute to contagion, we collected data from two social media sites: Digg and Twitter.  The microblogging service Twitter allows registered users to broadcast short messages, called tweets, to their followers. A message may contain a URL to external web content. In addition to posting a new message, a user can also retweet an existing message, analogous to forwarding an email. Twitter users create social links by following other users.  Each link is directed: we refer to the followed user as the friend, and the following user as the follower. Upon visiting Twitter, a user is presented with a stream containing tweets made by friends, ordered as a first-in last-out queue, with the most recent tweet (or retweet) at the top of the queue.

Social news aggregator Digg leverages opinions of its users to help people discover interesting news stories. Users submit URLs to news stories and vote for, or \emph{digg}, stories submitted by others. Users can follow the activity of others. The social user-interface on Digg shows a user a stream of stories his or her friends recently submitted or voted for. The stream is ordered chronologically by time of earliest recommendation (submission or vote) by a friend, with the most recent newly-recommended story at the top. When a user votes for a story, the recommendation is broadcast to a user's followers. However, additional recommendations do not change the story's relative position in the user's default social stream. Instead, a badge appears next to the story telling user how many friends have recommended it. When the story receives enough votes, Digg promotes it to the front page. However, before promotion, it can be found through friends' recommendations or on the newly submitted stories list, which at the time of data collection was receiving tens of thousands of new submissions daily.

We used Twitter's Gardenhose API to collect tweets over a period of three weeks in the Fall of 2010. We retained tweets that contained a URL in the body of the message. We used Twitter's search API to retrieve all tweets containing  those URLs, ensuring the complete tweeting history of all URLs, giving us more than 3 million tweets in total. We also collected the friend and follower information for all tweeting users, resulting in a social graph with almost 700K nodes and over 36M edges. We filtered out URLs whose retweeting behavior exhibits patterns associated with spam or automatic activity~\cite{Ghosh2011snakdd}, leaving us a data set containing 2K distinct URL's retweeted a total of 213K times. We use time stamps in tweet metadata combined with the follower graph to track when users are exposed to URLs by a friend and when they retweet them. We define a retweet to be anytime a user tweets a URL that had previously appeared in her Twitter feed. After removing spam URLs,  we only consider events where users received a particular URL less than 20 times, to further eliminate likely spam URLs.

We used the Digg API to collect data about 3.5K stories promoted to the front page in June 2009 and the times at which 140K distinct users voted for these stories. We also collected information about voters' friends, giving us a social graph with 280K users and 1.7M links.  For the present analysis, unless noted otherwise we consider only the voting dynamics occurring before promotion to the front page, so the primary means of information propagation is through the friends interface.  Both datasets were divided into training and test sets to rule-out over-fitting in determining the correct interpretation of the data.

\section{Results}
Using URLs as markers, we study the spread of information through the follower graphs of Digg and Twitter. A user may be exposed multiple times by friends to a URL. The exposure response function gives the probability of an infection as a function of the number of such exposures. An exposure occurs when a message containing the URL arrives in the user's stream, even if the user does not consciously see it. When aggregated over all users, both Twitter and Digg exposure response functions suggest complex contagion~\cite{Romero:2011um}:  while initial exposures increase infection probability, further exposures appear to saturate (Twitter) or suppress (Digg) further infection (Fig.~\ref{fig:exposureresponse}). Aggregated exposure response obscures heterogeneous behavior, because it conflates the response of users with different cognitive loads, i.e., different quantities of information in their stream. A large volume of incoming information, which scales with the number of friends a user follows as $n_f^{1.14}$,  reduces the user's ability to find any specific message~\cite{Hodas:2013tx,Hodas:2013we}. The likelihood a user will find a message containing the URL, therefore, has a normalization factor that depends on the number of friends, denoted $\mathcal{P}(n_f)$~\cite{Hodas2012}.

\begin{figure*}[hp]
\centering
   \subfigure[Twitter]{
   \includegraphics[width=3.6in]{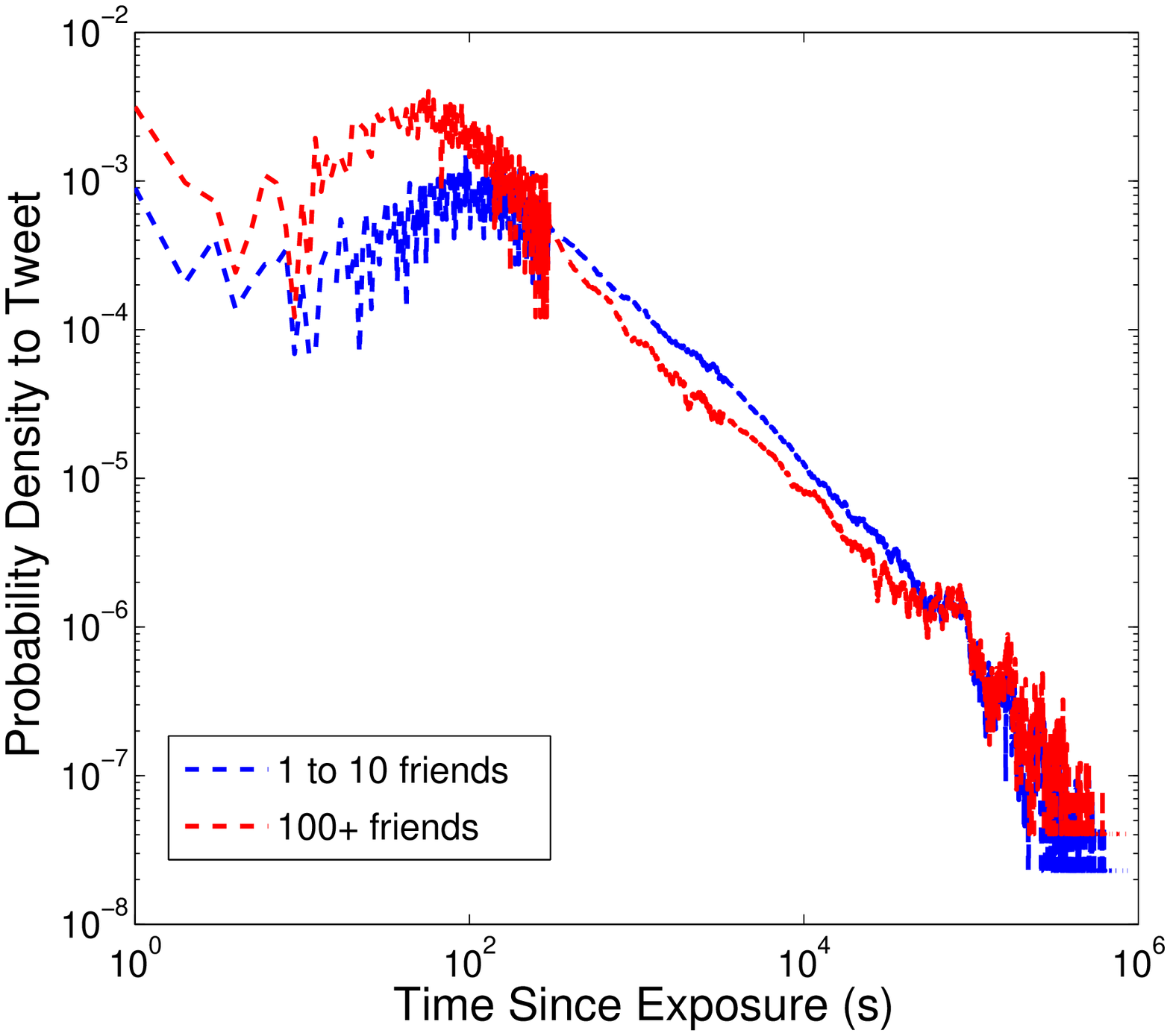} \label{fig:twittertimeresponse}
   }
   \subfigure[Digg]{
   \includegraphics[width=3.6in]{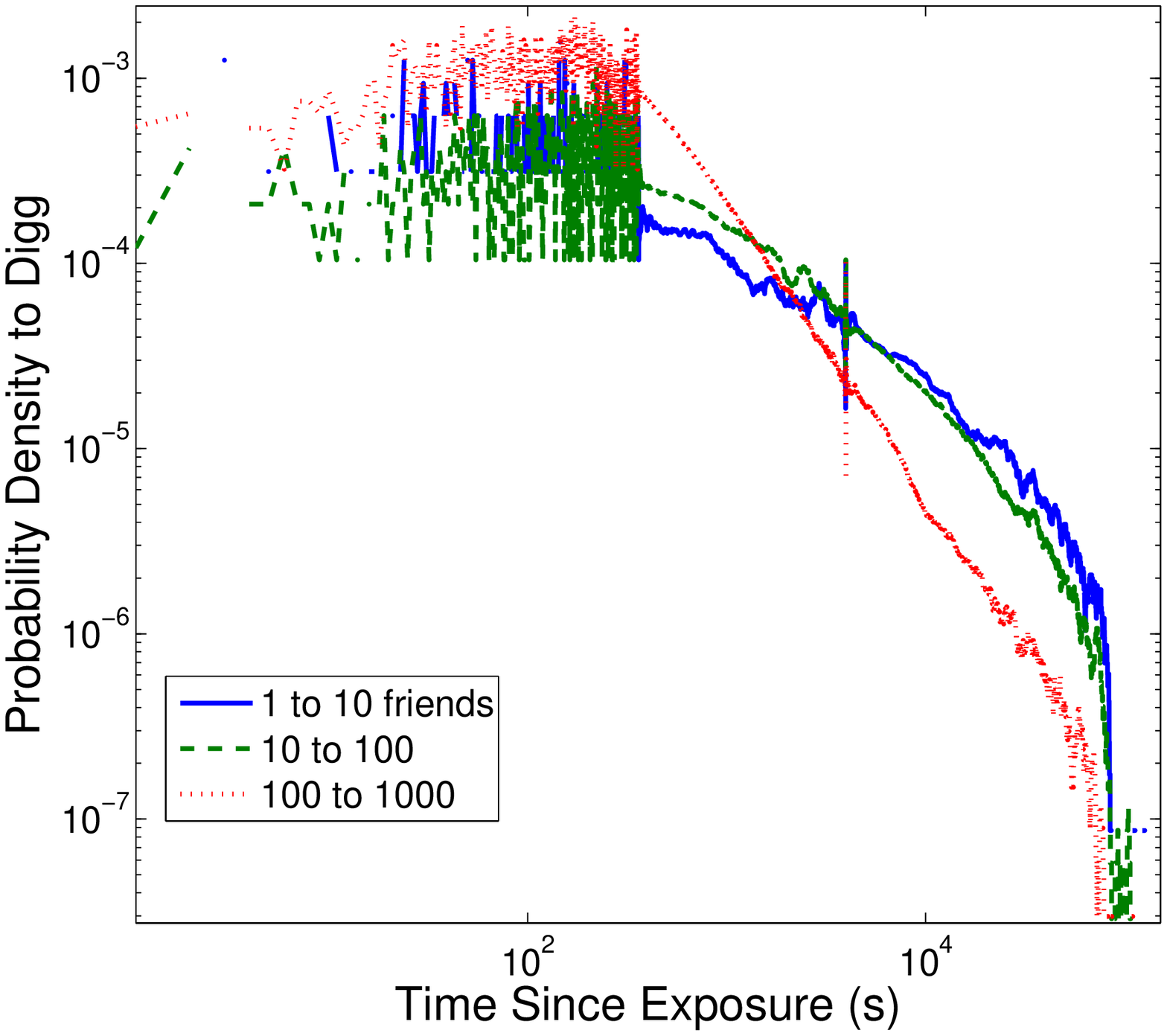}\label{fig:diggtimeresponse}
   }
   \caption{The time response function, the probability density of voting for a URL at a given time, drops off rapidly. Digg stories were only followed until promotion, which occurs at most 24 hours after appearing on Digg.  The data are smoothed using progressively wider smoothing windows.}
   \label{fig:timeresponse}
\end{figure*}

To become infected, a user must first discover at least one message containing the URL. The likelihood the user will see a specific message depends on its position in the user's stream. We use the term `visibility' to refer to this quantity. A new message starts at the top of the queue, where it is highly likely to be seen, because users usually start browsing from the top of a page~\cite{Buscher09}. With time, newer messages push it down the queue, where a user is less likely to see it before he or she gets bored, distracted, or leaves the site for any other reason~\cite{citeulike:5839088, citeulike:523633}. We measure a message's dynamic visibility using the  time response function, $\mathcal{T}(\Delta t, n_f)$, the probability that a user with  $n_f$ friends  retweets or votes at a time $\Delta t$  after the exposure~\cite{Hodas2012}. We plot  $\mathcal{T}(\Delta t, n_f)$  for Twitter and Digg in Fig.~\ref{fig:twittertimeresponse} and~\ref{fig:diggtimeresponse}, respectively, demonstrating that the visibility of a new message decays rapidly in time. Digg stories were only followed until promotion, which occurs at most 24 hours after appearing on Digg.  The data are smoothed using progressively wider smoothing windows, as in~\cite{Hodas2012}.	

A model describing user response to multiple exposures must consider the visibility of each exposure. In addition, a website's use of any social signals --- for example, displaying the number of friends who recommended the URL --- may alter user response, given that they have found the URL. The probability that a user with $n_f$  friends will be infected after $n_e$  exposures is
\begin{equation}\label{generalmodel}
P(t;n_e,n_f) = \sum_{n=1}^{n_e} F(n) V_n(t,\{t_1,\dots,t_{n_e}\};n_f),
\end{equation}
where $V_n()$ is the probability of finding $n$ of the $n_e$ exposures occurring at the times $t_1,\dots,t_{n_e}$, and $F(n)$ is the social enhancement factor accounting for the user observing that $n_e$ of their fiends have recommended the story.  Note that this formalism averages out content-specific factors and variable weights that a user may ascribe to different friends.

The particular functional form of $V_n$ depends on details of the website user-interface. On Twitter, all messages start at the top of the stream. By scanning the stream, a user can discover each message independently, so any of the exposures can result in an infection.  This behavior is well approximated by the probability of becoming infected by at least one exposure (see Supplement), given by
\begin{align}
P_{Twitter}(t;n_f,n_e) = &P_0 F(n_e) (1-\prod_{i=1}^{n_e}1-\mathcal{P}(n_f)\mathcal{T}(\Delta t_i, n_f)\nonumber\\
			 &+ v_{min},\label{twittermodel}
\end{align}
where $v_{min}$ is the effective minimum visibility of a message in the Twitter interface, the proportionality $P_0$ is fitted by minimizing weighted mean absolute percent (WMAP), as described in the Supplementary Methods, and $n_e$ is the number of exposures to the URL at time $t$. Underlying activity rates and cultural norms vary from site to site, so the proportionality $P_0$  can be interpreted as a task-specific scale factor.  The effective minimum visibility exists due to a user's ability to discover the URL outside the social media site or via other interfaces.

 We calculate  $\mathcal{P}(n_f)$ by measuring the average probability of retweeting the URL for users who were exposed once and only once to it. The average is taken over all users with  $n_f$ friends, as described in~\cite{Hodas2012,Hodas:2013we}.  The time response function  $\mathcal{T}(\Delta t_i, n_f)$ describes the visibility of a message since exposure at $t_i$. This is given by probability, shown in Fig.~\ref{fig:timeresponse}, that a user with $n_f$   friends will retweet a time   $\Delta t_i$ after the exposure, given that retweeting occurred. The time response function, $\mathcal{T}(\Delta t,n_f)$ is produced by calculating the probability that a user retweets/votes at the indicated interval  $\Delta t$ after a URL's arrival, given that the user votes on that URL.

The Digg user-interface differs from Twitter in that messages are by default ordered by the time of their \emph{first} appearance in the user's stream.  Any additional votes do not alter its position, but are reflected in a badge next to the URL that shows the number of friends, $n_e$, who voted for the URL.  The badge provides a social signal, which may alter user response.  Because of the user-interface, Eq.~\eqref{generalmodel} reduces to
 \begin{equation}\label{diggmodel}
 P_{Digg}(t;n_f,n_e) = F^\prime(n_e)\left(P^\prime_0 \mathcal{P}^\prime(n_f) \mathcal{T}^\prime(\Delta t,n_f) + v_{min}^\prime\right),
 \end{equation}
where $\Delta t$ is the time elapsed from the first vote by a friend, and the primes indicate Digg specific values for each quantity.  We empirically determined $F^\prime(n_e)$ using a maximum likelihood estimate, described in the Supplementary Methods. Social feedback in Digg results in large amplification of the probability of infection, shown in Fig.~\ref{fig:socialfeedbackdigg}. This could have multiple origins, including endorsement by friends~\cite{Bond:2012ff}, or from the increased visibility of the URL via alternative ways of discovering it on Digg, such as sorting URLs by popularity.

\begin{figure*}[hp]
\centering
\subfigure[Twitter, No Social Enhancement]{
\includegraphics[width=\columnwidth]{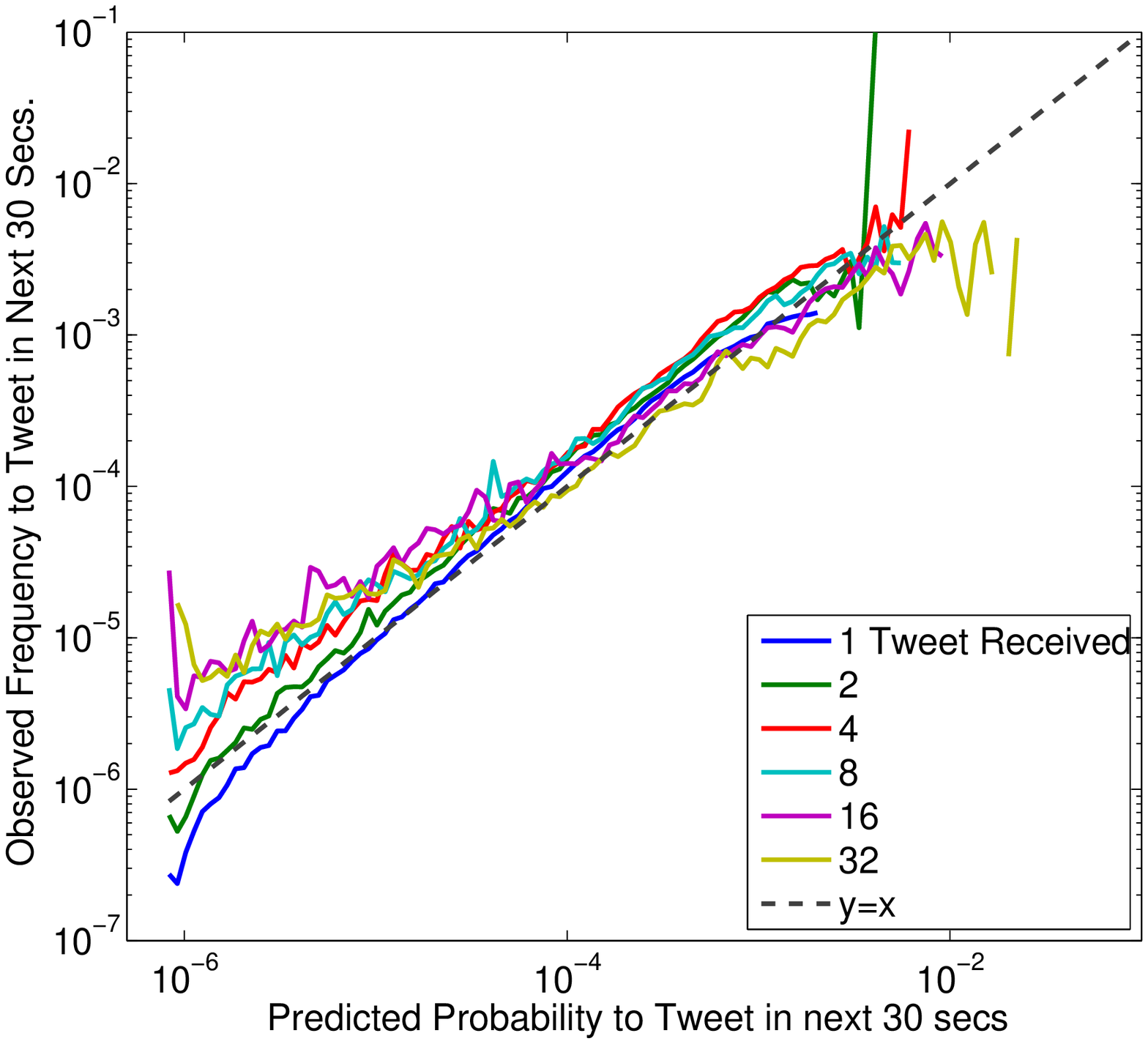}\label{fig:TwitterResultsNoSocial}
}
\subfigure[Twitter, Full Model]{
\includegraphics[width=\columnwidth]{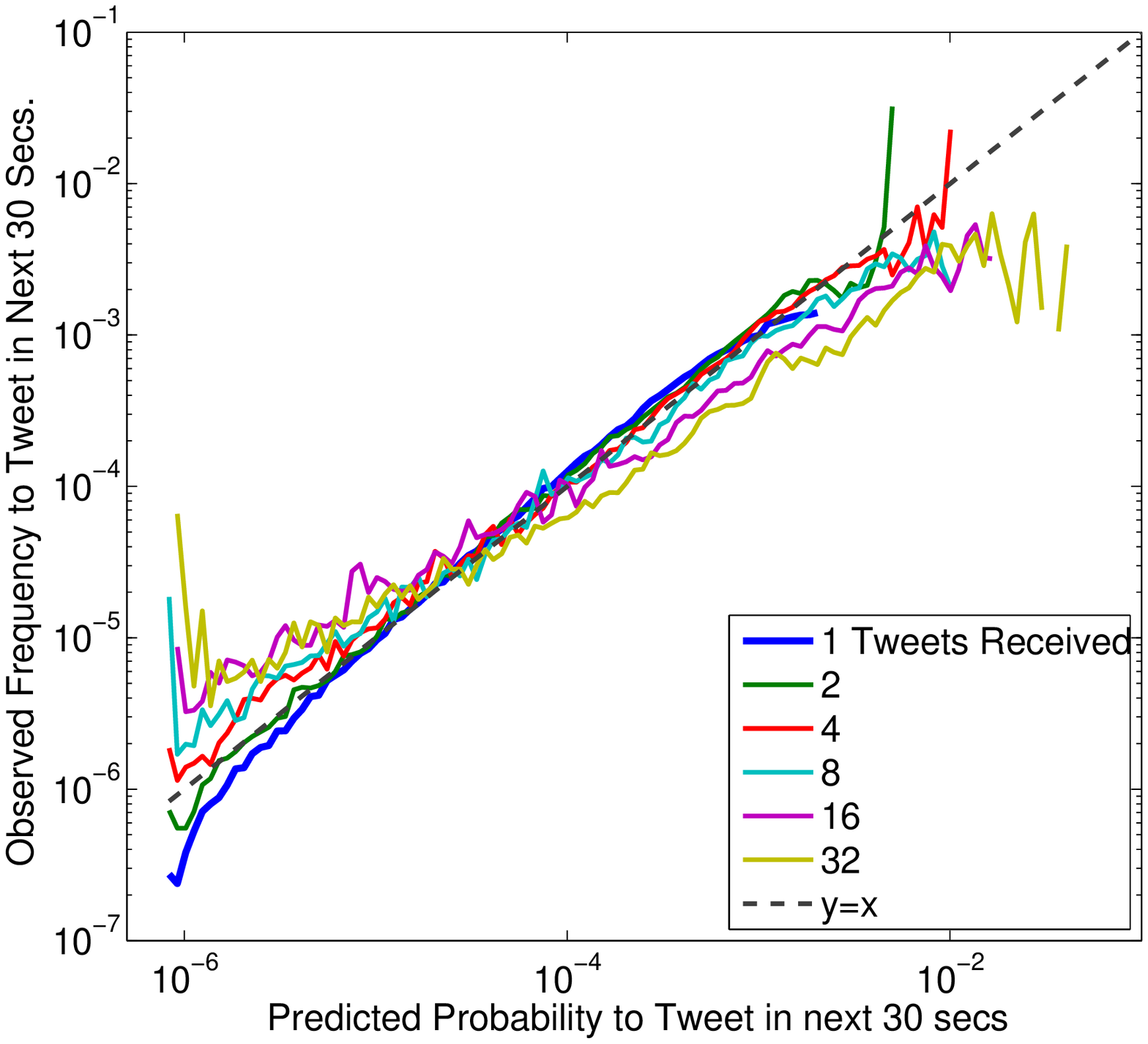}\label{fig:TwitterResultsFull}
}\\
\subfigure[Digg, No Social Enhancement]{
\includegraphics[width=\columnwidth]{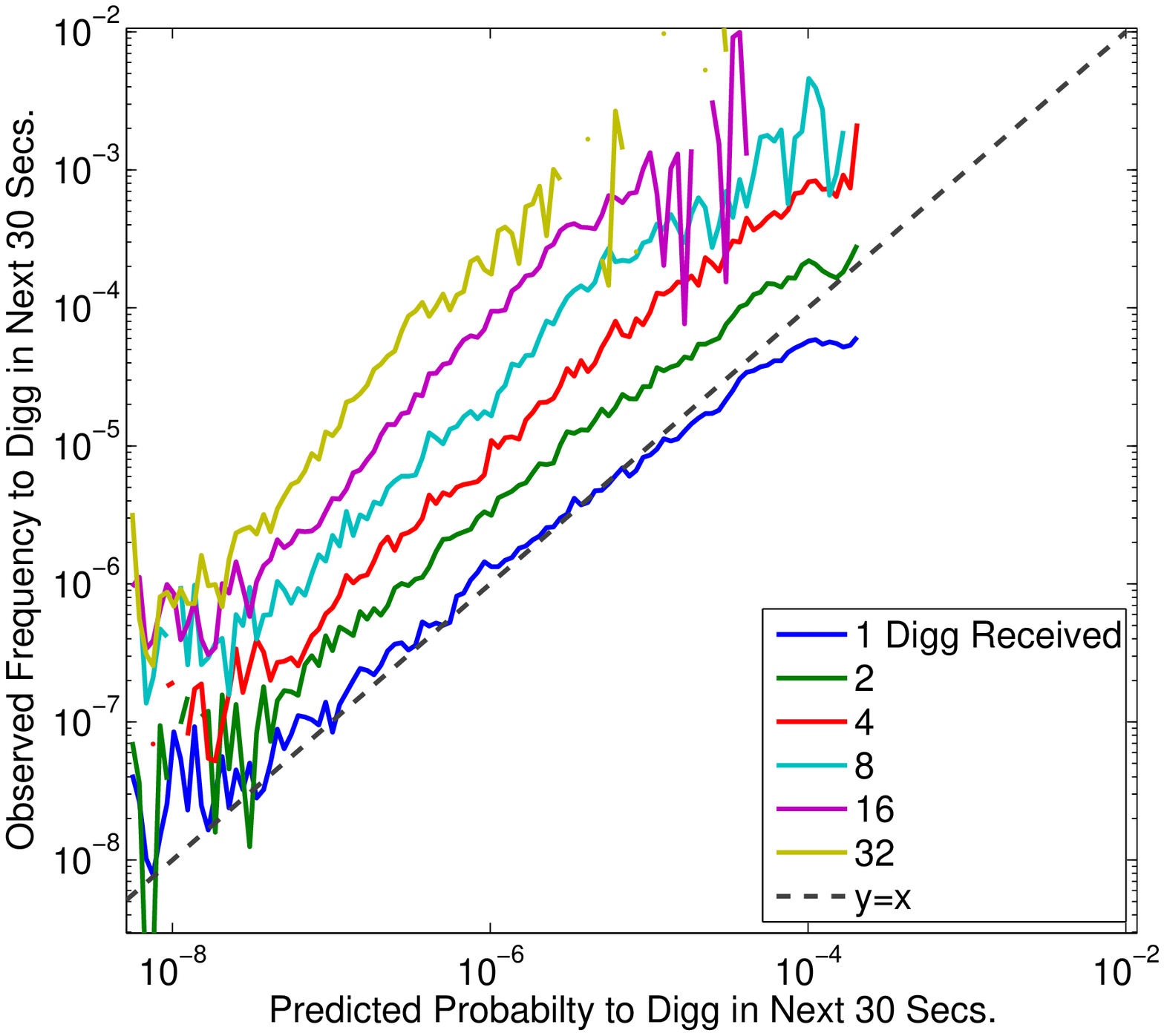}\label{fig:DiggResultsNoSocial}
}
\subfigure[Digg, Full Model]{
\includegraphics[width=\columnwidth]{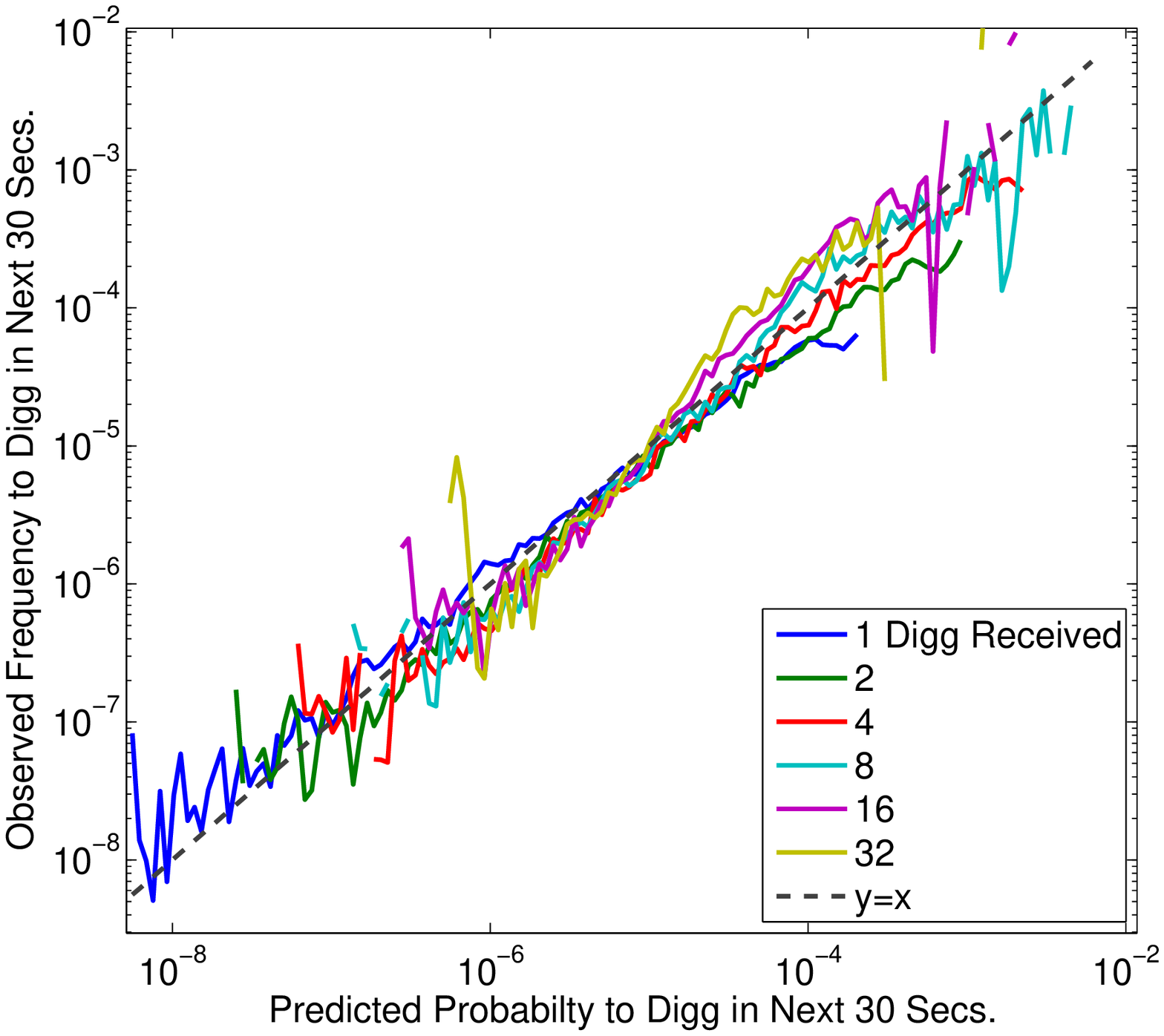}\label{fig:DiggResultsFull}
}
\caption{The social enhancement factors for Twitter and Digg. (A,C) Averaged over all users, (B,D) Calculated for sub-populations based on the number for friends, $n_f$. The decay in the social enhancement factor for Twitter can be attributed to residual spam in the dataset.\label{fig:ForecastResults}}
\end{figure*}

To validate the proposed model of social contagion, we forecast user activity and compare it to observed activity.  Specifically, we calculate  the observed frequency that a user with $n_f$  friends retweeted a URL in our Twitter dataset or voted for one in the Digg dataset in the subsequent 30 seconds.  Then, using Eq.~\eqref{twittermodel} or Eq.~\eqref{diggmodel}, we calculate the theoretical probability that a user with  that many friends would act in those 30 seconds, given the same exposures.   Plotting the predicted versus observed probabilities allows us to graphically assess the accuracy of the contagion model.  Unbiased forecasts lie along the  line of the graph. The proposed model accurately forecasts response to multiple exposures on Twitter (Fig.~\ref{fig:TwitterResultsFull}) and Digg (Fig.~\ref{fig:DiggResultsFull}), indicated by a WMAP error of 0.5\% and 1.5\%, respectively.  Ignoring social enhancement, and thereby utilizing a model akin to an ICM, produces systematically biased results, shown in Figs.~\ref{fig:TwitterResultsNoSocial} and~\ref{fig:DiggResultsNoSocial}.  Hence, any ICM-based model, such as a traditional SIR, could not achieve unbiased forecasting without accounting for social enhancement. Similarly, a model that does not account for visibility decay could not account for  variations in user-interface, evident in the difference between Eqs.~\eqref{twittermodel} and~\eqref{diggmodel}.

\remove{
}

Rapid decay of visibility, combined with decreased susceptibility of highly connected users, explains why information in social media fails to spread as widely as predicted by the generic ICM~\cite{VerSteeg11}. We can rule out novelty decay, at least on Twitter, because after taking visibility into account, infection probability does not depend on the age of the information~\cite{Hodas2012} We cannot evaluate novelty decay for Digg, because we examine only the votes URLs receive within 24 hours of submission (before it is promoted to the front page), which is too short a time period to see novelty decay. Although different types of information may spread according to slightly different patterns~\cite{Yang:2011wt,Wu:2011vs} our analysis is content agnostic, so the reported results are the population average, with the caveat that we have removed most spam.
Explicit social feedback can significantly magnify user response, albeit making it less useful for popularizing high-quality content~\cite{Salganik06}. Unlike Digg, the Twitter user-interface offered no explicit social feedback (beyond trending topics). Users may remember seeing a friend's recommendation of the URL, a factor that could explain the slight social enhancement seen in Twitter response in Fig.~\ref{fig:socialfeedbacktwitter}.  When explicit social feedback is present, as in Digg, Fig.~\ref{fig:socialfeedbackdiggbreakdown} shows that users appear to  weigh their actions based on the fraction of friends endorsing a URL instead of considering the absolute number.
This effect could explain complex contagion, in which ``network effects" appear to play a significant role in the contagion process~\cite{Centola:2010ww}.

\section{Conclusions}
\begin{figure*}[hp]
\centering
\includegraphics[width=3.6in]{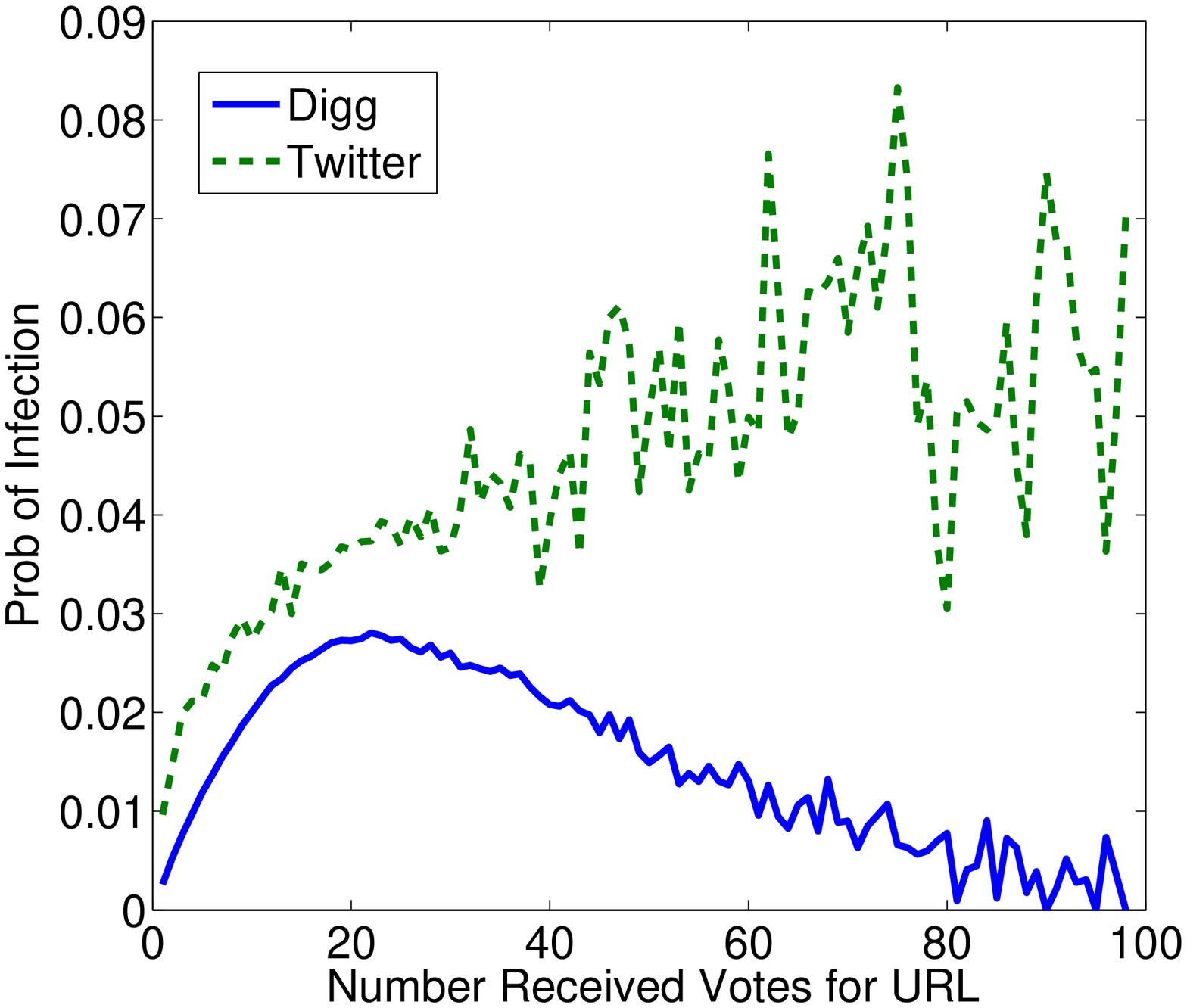}
\caption{The exposure response function for Digg and Twitter averaged over all users.  In Digg, a vote is a digg by a friend. In Twitter a vote is a received tweet containing the URL. This averaging gives the appearance of reduced susceptibility to repeated exposure.}\label{fig:exposureresponse}
\end{figure*}

\begin{figure*}[hp]
\begin{center}
\subfigure[Twitter, All Users]{
\includegraphics[width=\columnwidth]{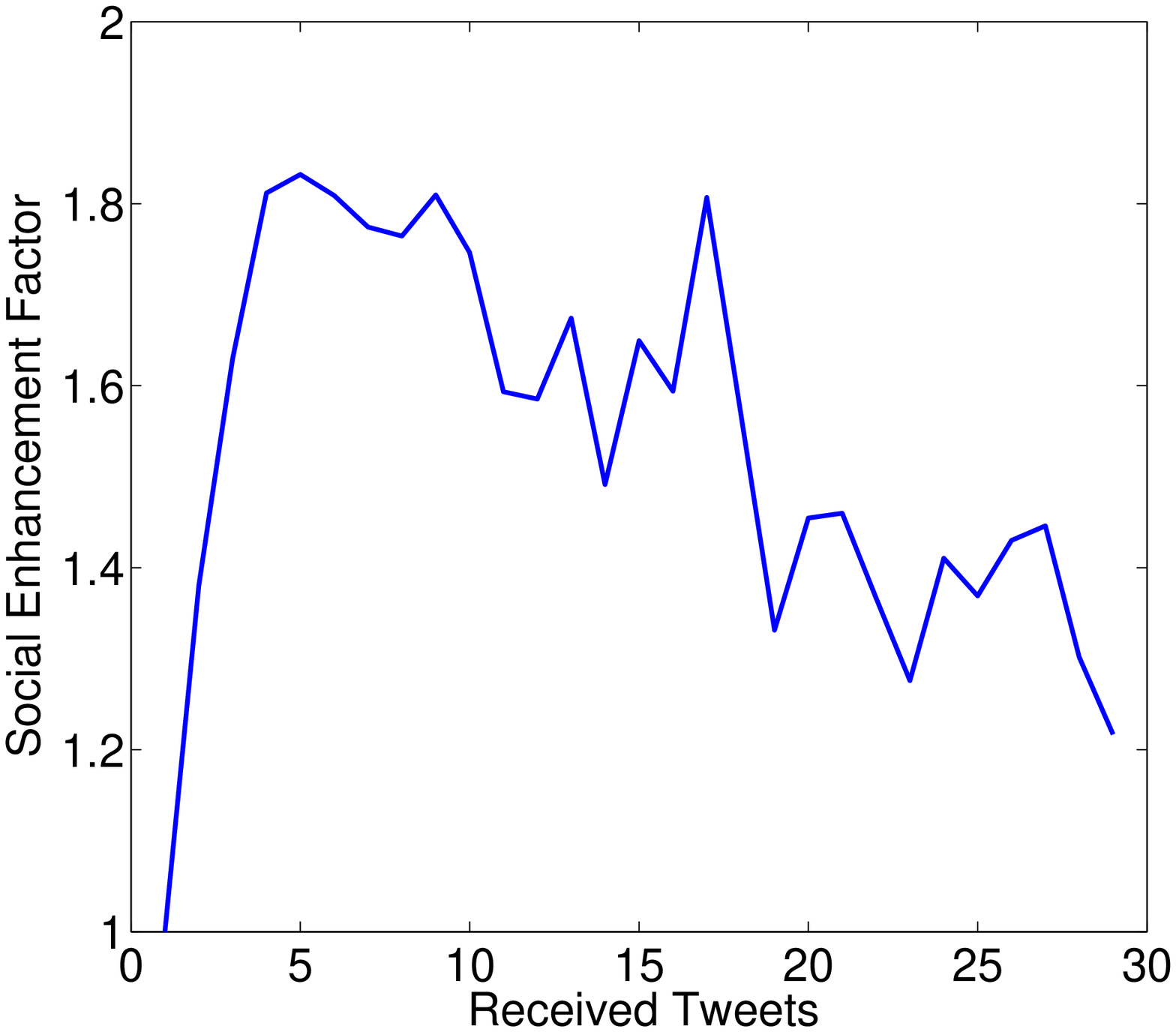}\label{fig:socialfeedbacktwitter}
}
\subfigure[Twitter, Sub-populations]{
\includegraphics[width=\columnwidth]{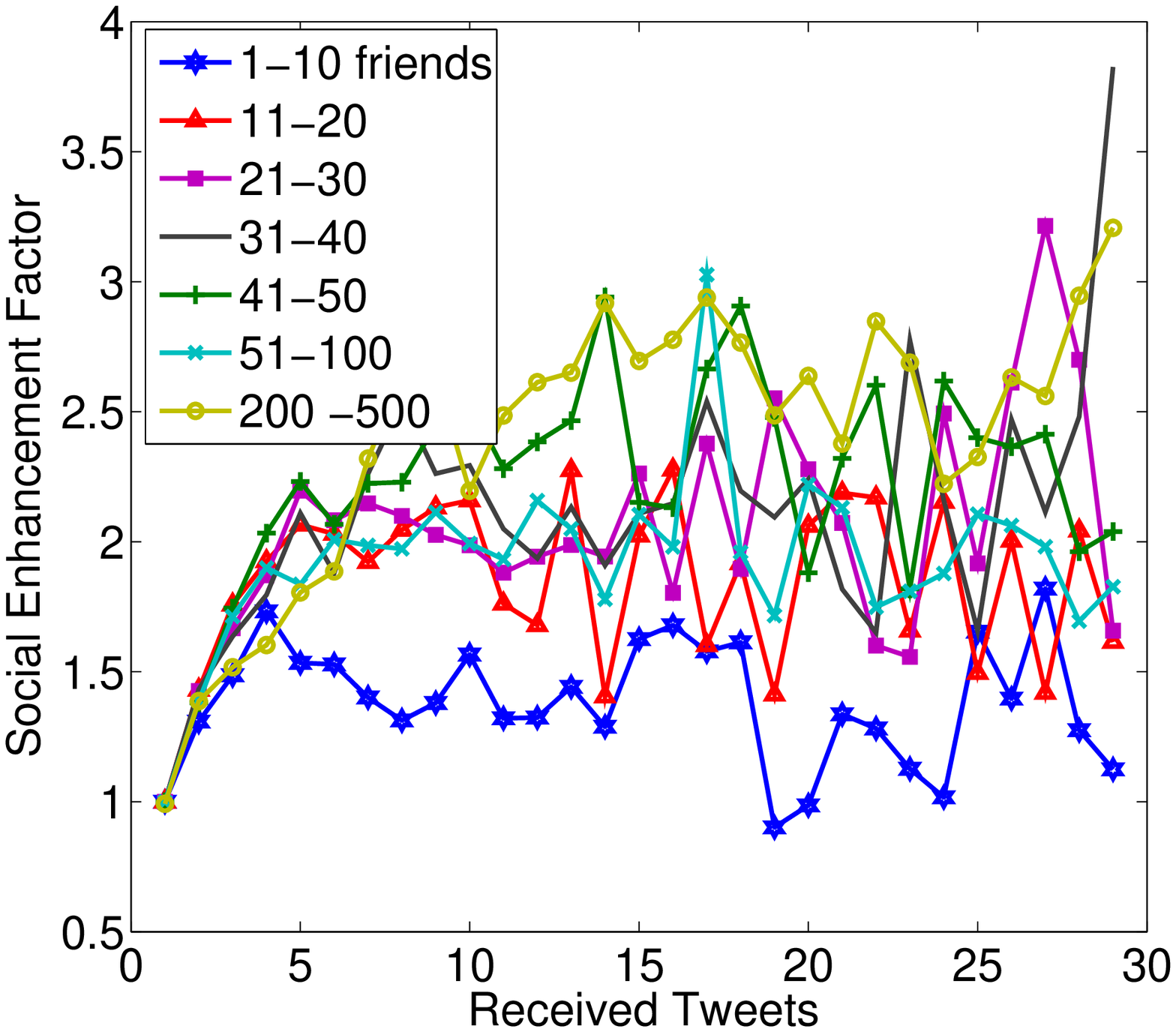}\label{fig:socialfeedbacktwitterbreakdown}
}\\
\subfigure[Digg All Users]{
\includegraphics[width=\columnwidth]{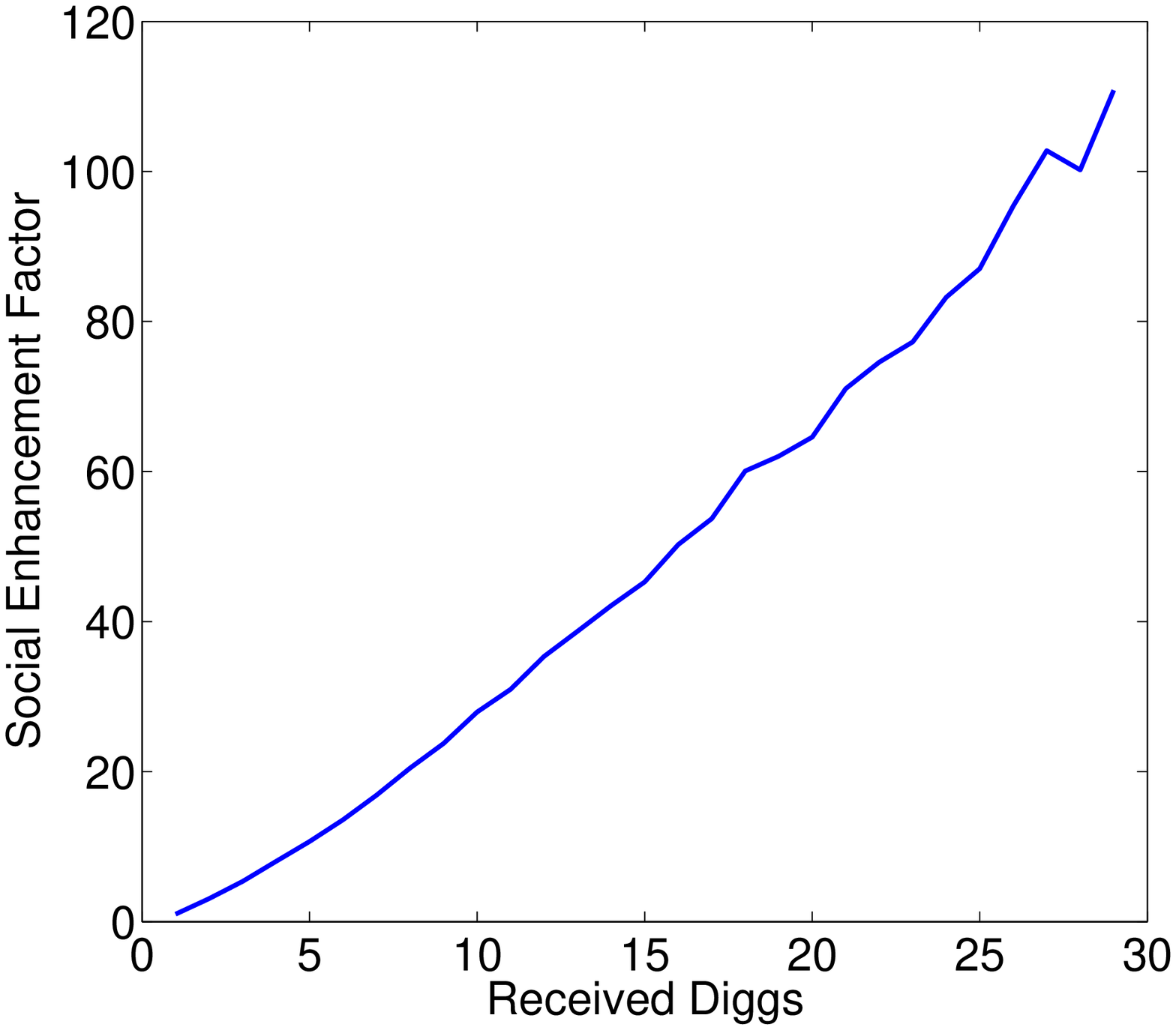}\label{fig:socialfeedbackdigg}
}
\subfigure[Digg, Sub-populations]{
\includegraphics[width=\columnwidth]{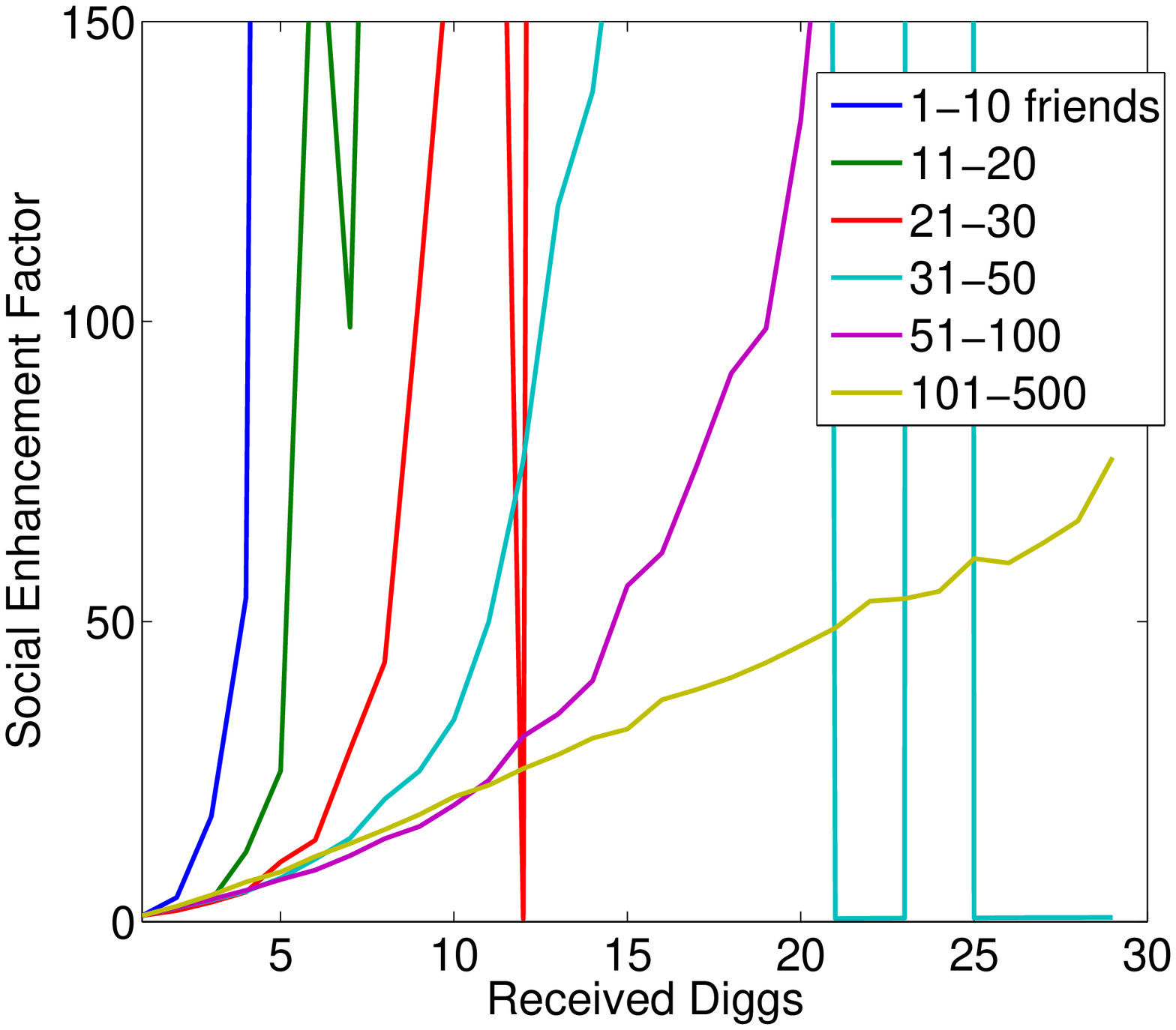}\label{fig:socialfeedbackdiggbreakdown}
}
\end{center}
\caption{The social enhancement factors for Twitter and Digg. (A,C) Averaged over all users, (B,D) Calculated for sub-populations based on the number for friends, $n_f$. The decay in the social enhancement factor for Twitter can be attributed to residual spam in the dataset.\label{fig:social feedback}}
\end{figure*}

The present results show that there are important and surprising differences between the spread of information and a disease that stem from human cognitive limitations for processing information. In pathogenic contagion, highly connected people amplify the spread of disease, but in social contagion they inhibit the spread. This is because these people are so overloaded with other information their friends post, they are less likely to notice and act on a particular piece of information, and they require stronger social signal to act (Fig.~\ref{fig:socialfeedbackdiggbreakdown}), on average.
As the visibility of a message decays, users are less likely to expend the effort required to find it. Because the volume of information scales with the number of friends a user follows, visibility decays faster for users with more friends, making highly connected users far less susceptible to any single exposure than poorly connected users~\cite{Hodas2012,Hodas:2013we}.  Users with many friends are less likely to respond, and they dominate the high-exposure portion of the average exposure response function (Fig.~\ref{fig:exposureresponse}), giving the impression that more exposures may be counter-productive.  On the contrary, the present work suggests that people are more susceptible to repeated exposure than the population average suggests.

By comparing two different websites with very different user-interfaces, we have demonstrated that it is possible to isolate the factors in social contagion due to social feedback and the user-interface, without directly manipulating the underlying social network or user-interface~\cite{Bakshy:1418056,Bond:2012ff,Centola:2010ww}. Moreover, the unbiased fidelity of the proposed model suggests that once visibility of the exposures is taken into account, social contagion operates as a simple contagion, i.e., with infection probability increasing monotonically with the number of exposures, which can be quantified by utilizing properly normalized time response functions.

Our work highlights how cognitive constraints impact information processing in everyday activities.  While humans have developed large brains, partly to handle the mental demands of social life~\cite{Dunbar03,Silk07}, cognitive constraints imposed by our brain's finite capacity to process information affects social behavior, for example, by limiting maximum group size~\cite{Dunbar}.  Cognitive constraints also affect how individuals utilize their dynamic information streams in social media. Attentive acts, such as browsing a website and reading tweets, require mental effort, and since the brain's capacity for mental effort is limited by its energy requirements, so it attention~\cite{Kahneman73}.
\remove{
}
This will tend to reduce the likelihood of response under conditions of high information load. Thus, social contagion will be highly dependent on explicit social feedback and the user-interface.

Regardless of the social synergy desired by the social network, visibility of the URL appears to be an essential factor in determining accuracy of activity forecasts.   Because users only dedicate a limited amount of time and effort interacting with any website, the site's visibility policy will largely determine the quality of the user experience with respect to information discovery and propagation. Thus, because Digg does not refresh the position of a URL after each recommendation, the social signals it uses do not compensate for the loss of visibility the URL suffers over time. Each website's design choice can broaden or narrow the user-base's attention, but whatever information occupies a position of high visibility will dominate social contagion originating from that site.  Although the current work provides techniques for real-time forecasting of the average user behavior on a specific website, understanding the emergence of globe-spanning viral content will require accounting for the interaction of the dynamic visibility and social synergy across a multitude of websites and media outlets.

\begin{acknowledgments}
-- text of acknowledgments here, including grant info --
\end{acknowledgments}
\bibliographystyle{abbrv}
\bibliography{biblibrary_standard}








\pagebreak
\section{Supplementary Material}
\subsection{Methods}
To calculate probabilities of response to multiple exposures, the data was broken down into separate time series, each corresponding to the arrival of specific URL-containing tweets or votes into a single user's stream. For each series, at every one-second interval we calculate the quantity we define as `visibility' of the URL:
\begin{align*}
	V_{all}(t,n_f) &= 1 - \prod_d (1-\mathcal{T}(t-t_i,n_f)),\text{ or}\\
	V_{first}(t,n_f) &= \mathcal{T}(t-t_0,n_f),
\end{align*}
where $n_f$ is the number of friends of the user, $\mathcal{T}(\Delta t,n_f)$ is the time response function for a user with $n_f$ friends. $V_{all}$  is  proportional to the probability of finding any one of the received messages at time $t$ , while  $V_{first}$ is proportional to the probability of finding only the first  message.

 We calculate  $\mathcal{P}(n_f)$ by measuring the average probability of retweeting the URL for users who were exposed once and only once to it. The average is taken over all users with  $n_f$ friends, as described in~\cite{Hodas2012,Hodas:2013we}.  The time response function  $\mathcal{T}(\Delta t_i, n_f)$ describes the visibility of a message since exposure at $t_i$. This is given by probability, shown in Fig.~\ref{fig:timeresponse}, that a user with $n_f$   friends will retweet a time   $\Delta t_i$ after the exposure, given that retweeting occurred.

The time response function, $\mathcal{T}(\Delta t,n_f)$ is produced by calculating the probability that a user retweets/votes at the indicated interval  $\Delta t$ after a URL's arrival, given that the user votes on that URL. For Twitter data we calculate the time response function only for those events in which a user received the URL once and only once. For Digg, this constraint is lifted, because there are too few such events in the Digg data. The precise time response function depends on $n_f$ , because users with many friends receive new messages at a higher rate, causing the visibility of any specific message to decay more quickly~\cite{Hodas2012}.   We lack sufficient data to precisely calculate the time response function for each  $n_f$.  Instead, we calculated the time response function for users with $n_f = 1-2$, $n_f=9-11$ , and  $n_f = 90-110$, producing $\mathcal{T}_1$, $\mathcal{T}_{10}$, and $\mathcal{T}_{100}$, respectively, following the procedure in~\cite{Hodas2012}.  To estimate the time response function for arbitrary $n_f$, we interpolated as follows:
\begin{align*}
	w_1 &\equiv ((n_f-1)^2 + 10^{-6})^{-1}\\
	w_{10} &\equiv ((n_f-10)^2 + 10^{-6})^{-1}\\
	w_{100} & \equiv ((n_f-100)^2 + 10^{-6})^{-1} \text{ for Twitter, or}\\
	w_{100} & \equiv (|n_f-100|+ 10^{-6})^{-1} \text{ for Digg}\\
	\mathcal{T}(\Delta t,n_f) & =\frac{ w_1 \mathcal{T}_1(\Delta t) + w_{10} \mathcal{T}_{10}(\Delta t)+ w_{100} \mathcal{T}_{100}(\Delta t)}{w_1+w_{10}+w_{100}}.
\end{align*}

To produce the fits for $v_{min}$ and $P_0$, we plot the calculated probability versus the observed probability for an event, i.e. forming a function $O(p)$, where $p$ is the calculated probability.  We isolated the events corresponding to a receiving a single message, leading to a subset of predictions denoted $O_1(p)$. We then minimize the weighted mean absolute percent error (WMAP)~\cite{Armstrong:1992uj},
\begin{equation*}
 \left \langle \left | P_0 O_1(p) + v_{min} - p\right| / p\right\rangle
 \end{equation*}
by searching over $P_0$ and $v_{min}$.  For Digg, we have $P_0 = 667$, $\log(v_{min}) = -19$.  An analytical form for $\mathcal{P}(n_f)$ was determined by fitting to minimize RMS error of the empirically determined $\mathcal{P}(n_f)$~\cite{Hodas2012}, giving Digg's $\mathcal{P}^\prime (n_f) = A/((e^{B n_f} + C)(n_f +D)(n_f+E))$, where $A=7.6\cdot10^{-3},B=-6.2\cdot10^{-2},C=1.7\cdot10^{-3},D=3.7,E=17.8$\footnote{Note the $E$ was chosen by minimizing $WMAP$ error simultaneously with fitting $P_0$ and $v_{min}$ on the training data, i.e. $E$'s purpose is to correct for sparsity in the empirically calculated $\mathcal{P}(n_f)$ for Digg.}. For Twitter we have $P_0=16.6$ and $\log(v_{min}) = -14$, and we used $\mathcal{P}(n_f) = A n_f^P/(n_f+B)$, where $A=0.3,P=0.16,C=0.55$.

To calculate the social enhancement factors, we carry out the MLE for $F(n_e)$ in the following manner. We take as axiomatic the true probability of a response given  $n_e$ exposures is $F(n_e)P(\nu)$, where $\nu$ parameterizes the underlying visibility.  Thus, given $N(\nu)$  observed events for a specific  $\nu$, the likelihood, $\ell$, of observing $N_{r}(\nu)$  responses is determined by the binomial distribution
\(\ell(\nu,n_e) = \left(\begin{array}{c}N(\nu) \\N_{r}(\nu)\end{array}\right) \left(F(n_e)P(\nu)\right)^{N_r(\nu)} \left(1-F(n_e)P(\nu)\right)^{N(\nu)-N_r(\nu)}.\)
The total log-likelihood of observing the curve $O_{n_e}(\nu)$ is thus
\begin{align*}
\mathcal{L}(n_e) = \sum_{\nu} & \log \left(\begin{array}{c}N(\nu) \\N_{r}(\nu)\end{array}\right) \\
			+& N_r(\nu)\left(\log F(n_e) + \log P(\nu)\right)\\
			+& \left(N(\nu)-N_r(\nu)\right)\log \left(1-F(n_e)P(\nu)\right).
\end{align*}
For each value of $n_e$, we find the value of $F(n_e)$ that maximizes $\mathcal{L}(n_e)$. First, for $n_e=1$, we define $F(1) = 1$, so we obtain the MLE for $P(\nu)$ using
\begin{align*}
\frac{\partial}{\partial P(\nu)} \mathcal{L}(1) = &\frac{N_r(\nu)}{P(\nu)} - \frac{N(\nu)-N_r(\nu)}{1-P(\nu)} = 0,
\end{align*}
giving $P(\nu)=N_r(\nu)/N(\nu)$. Then, for $n_e>1$, we are left to find the likelihood maximizing $F(n_e)$ given $P(\nu)$, leading to
\begin{align*}
\frac{\partial}{\partial F(n_e)} \mathcal{L}(n_e) = &\sum_\nu \frac{N_r(\nu)}{F(n_e)} - \left(N(\nu) - N_r(\nu)\right) \frac{P(\nu)}{1-F(n_e)P(\nu)}. 
\end{align*}
Numerically solving for $\frac{\partial}{\partial F(n_e)} \mathcal{L}(n_e) = 0$ provides the MLE for $F(n_e)$.

The minimum possible observed probability is bounded by the number of observed events.  In the forecasting predictions, the friend-cohort breakdown in Fig.~\ref{fig:ForecastResults} appears to deviate from the observed probabilities at very high and low predicted probabilities.  However, this is due to the minimum probability floor rising beyond the predicted=observed line, because events with high visibility and high social influence or very low visibility are less common.

\subsection{Approximating Visibility Functions}
Depending on the user-interface, the user may be exposed to a URL from a variety of different messages.  Under general conditions, the probability of discovering a URL will depend on the visibility of each of those messages.  In addition, the user's response may differ depending on how many times they actually observed the URL.   Thus, the probability of being infected by a URL will depend on the probability of seeing the URL  $n$  times and an enhancement factor, $f(n;n_f)$, arising from the collective effect of multiple exposures.  The probability of acting at time  $t$ is, therefore,
\[P(t,n_e;n_f) = \sum_{n=1}^{n_e} f(n;n_f)V_{n}(t,\{t_1,\dots,t_{n_e}\};n_f),\]
where $V_{n}(t,\{t_1,\dots,t_n\};n_f)$ is the probability of explicitly observing $n$ of the $n_e$ URL's that arrived at times $t_1, \dots,\ t_{n_e}.$ For Digg before promotion to the front page, the URL is ordered by the time of its first recommendation to the user, so $V_n(t) = \delta_{n,n_e} \mathcal{P}(n_f)\mathcal{T}(n,n_f)$, where $\delta_{n,n_e}$ is the Kronecker delta function.  Thus, only one term is relevant for Digg.  Approximating $f(n_e;n_f)$ as the social enhancement factor $F(n_e)$ gives the probability of Digging a URL a user receives to be Eq.~\eqref{diggmodel}.

For Twitter, each tweet is displayed based on the chronological order of its arrival, so there may be multiple tweets potentially containing the same URL in the user's stream.  Each tweet decays in visibility according to the time-response function, based on the time of its arrival in the user's stream.  Thus, the probability of discovering tweet $i$  containing the URL arriving at time $t_i$ is $\mathcal{P}(n_f)\mathcal{T}(t-t_i,n_f)$. For brevity, we will abbreviate this quantity as $\tau_i \equiv \mathcal{P}(n_f)\mathcal{T}(t-t_i,n_f)$. The probability of seeing a URL only once is
\[
V_1(t) = \sum_{i}^{n_e}\prod_{j\neq i}^{n_e} \tau_i (1-\tau_j).
\]
Similarly, the probability of seeing exactly two out of $n_e$ URL's is
 \[
V_2(t) = \sum_{i}^{n_e-1}\sum_{j>i}^{n_e}\prod_{j\neq i}^{n_e}\prod_{k\neq i,j}^{n_e} \tau_i \tau_j (1-\tau_k).
\]
As $n_e$ grows, the number of combinations required to enumerate each $V_n$ grows rapidly with $n_e$. $V_n$ will have $n_e!/n!(n_e-n)!$ terms. Although one could calculate all $V_n$ explicitly every time-step for every user and URL, this is currently computationally prohibitive.  We propose an approximate form for $P_{Twitter}$, Eq.~\eqref{twittermodel}, justified as follows.

Although each $V_n$ for Twitter may have many terms, it can be represented succinctly using a generation function,
\[
V_n = \mathbb{C}_{n_e} \frac{1}{n!}\frac{\partial^n}{\partial y^n}\prod_{i=1}^{n_e} \left(1+\frac{\tau_i}{1-\tau_i} y\right)\bigg|_{y=0},
\]
where $ \mathbb{C}_{n_e} \equiv \prod_{i=1}^{n_e}(1-\tau_i)$ is the probability of not seeing any of the $n_e$ tweets. Using this form, the exact expression for $P_{Twitter}$ is
\[
P_{exact} = \mathbb{C}_{n_e} \sum_{n=1}^{n_e} \frac{f(n;n_f)}{n!} \frac{\partial^n}{\partial y^n} \prod_{i=1}^{n_e}\left(1+\frac{\tau_i}{1-\tau_i} y\right)\bigg|_{y=0}.
\]
We wish to know how well $P_{exact}$ can be approximated if we take $f(n;n_f) = F_{tw}(n_e)$, i.e. determining the probability of seeing any URL, with a social enhancement factor.  Using generating functions, this approximation is expressed as
\begin{align*}
P^* =& \mathbb{C}_{n_e} F_{tw}(n_e)\sum_{n=1}^{n_e} \frac{1}{n!} \frac{\partial^n}{\partial y^n} \prod_{i=1}^{n_e}\left(1+\frac{\tau_i}{1-\tau_i} y\right)\bigg|_{y=0}\\
	=&\mathbb{C}_{n_e} F_{tw}(n_e)\left(e^{\frac{\partial}{\partial y} }- 1\right)\prod_{i=1}^{n_e}\left(1+\frac{\tau_i}{1-\tau_i} y\right)\bigg|_{y=0},
\end{align*}
where $e^{\frac{\partial}{\partial y}} = \sum_{n=0}\frac{1}{n!}\frac{\partial^n}{\partial y^n}.$ One may show that $e^{a\frac{\partial}{\partial y}} f(y) = f(y+a)$ by left-multiplying both sides of this identity by the inverse operator $e^{-a\frac{\partial}{\partial y}}$.  Using this identity gives
\begin{align*}
P^* =& \mathbb{C}_{n_e} F_{tw}(n_e) \left(\prod_{i=1}^{n_e} (1+\frac{\tau_i}{1-\tau_i})-1\right)\\
	=& F_{tw}(n_e) \left(1-\prod_{i=1}^{n_e}\left(1-\mathcal{P}(n_f) \mathcal{T}(\Delta t_i,n_f)\right)\right),
\end{align*}
where we have expanded all of the definitions in the last line above.  To find the best choice for $F_{tw}(n_e)$ to determine its suitability as an approximation, we define the ratio
\[
F^*(n_e,t) \equiv \frac{P_{exact}}{P^* / F_{tw}(n_e)} = \frac{\hat{S}_f G(y)}{\left(e^{\frac{\partial}{\partial y}}-1\right) G(y)}\bigg|_{y=0},
\]
where $\hat{S}_f \equiv \sum_{n=1}^{n_e} \frac{f(n;n_f)}{n!}\frac{\partial^n}{\partial y^n}$ and $G(y) \equiv \prod_{i=1}^{n_e}\left(1+\frac{\tau_i}{1-\tau_i}y\right)$.  That is, $F^*(n_e,t)$ is simply the time-dependent ratio of the exact expression for seeing one of $n_e$ tweets to the approximated form (without $F_{tw}(n_e)$).  If this quantity varies very little with time, then taking $F^*(n_e,t) \sim F_{tw}(n_e)$ will give a good approximation.
	
Because of the nature of the Digg interface, as stated above, we can directly observe plausible forms for the $f(n;n_f)$ enhancements.  We observe that Digg enhancements are generally linear, and we may surmise an approximate form for the Twitter enhancements to be $f(n;n_f) \approx \alpha n + \beta$. This is not to hypothesize a true form of $f(n)$ for Twitter but to merely provide a plausible function form to test the accuracy of the proposed approximation.  This gives
\begin{align*}
 \hat{S}_f =& \alpha \sum_{n=1}^{n_e} \frac{n}{n!} \frac{\partial^n}{\partial y^n} + \beta\left(e^{\frac{\partial}{\partial y}}-1\right) \\
 		=& \alpha \frac{\partial}{\partial y} e^{ \frac{\partial}{\partial y}} + \beta\left(e^{\frac{\partial}{\partial y}}-1\right).
\end{align*}
Replacing the first sum is possible because derivative-orders greater than $n_e$ evaluate to 0.  Returning to the expression for $F^*(n_e,t)$, we have
\begin{align*}
F^*(n_e,t) &= \frac{\alpha G^\prime(1) + \beta(G(1)-1)}{G(1)-1} \\
		&= \alpha \frac{\sum_{i=1}^{n_3} \tau_i}{1-\prod_{i=1}^{n_e} (1-\tau_i)} + \beta.
\end{align*}
Because each $\tau$ is proportional to the time-response function, $F^*_{tw}$ varies with time.  Because the probabilities of conducting any action on Twitter at any instant are low, $\tau \ll 1$.  Consider the two extreme, yet plausible, scenarios: 1) The user receives $n_e$ messages  simultaneously all with maximum visibility or 2) The user receives $n_e$ messages which have decayed to extremely low visibility, so $\tau\rightarrow v_{min}$. For case 1, the maximum visibility corresponds to $\mathcal{P}(n_f) \mathcal{T}(\sim0,n_f)$, which is very small, i.e., $<10^-3$~\cite{Hodas2012}.  For case 2, the minimum visibility is $v_{min}$, so for either limit we have
\begin{align*}
F^* = \alpha n_e \tau_{0} / \left(1-(1-\tau_{0})^{n_e}\right) + \beta \approx \alpha + \beta + \frac{\alpha}{2}  (n_e-1)\tau_{0}.
\end{align*}
In case 1, $\tau_0 = \tau_{max}$, and in case 2, $\tau_0 = v_{min}$.  Thus, in either limit, $F^*$ will tend to have a characteristic value of $\alpha + \beta$,  varying weakly with time, confirming the argument that the full $P_{exact}$ can be approximated by $P^*$.

\end{document}